\documentclass[aps,prb,twocolumn,showpacs,groupedaddress]{revtex4}
\usepackage{graphicx,amssymb,amsmath}
\usepackage{natbib}
\usepackage{bm}
\newcommand\FeCuGeO{Fe$_2$Cu$_2$Ge$_4$O$_{13}$~}

\newcommand\CuFeGeO{Cu$_2$Fe$_2$Ge$_4$O$_{13}$}
\newcommand\CuScGeO{Cu$_2$Sc$_2$Ge$_4$O$_{13}$}

\begin{document}
\title{Magnetic excitations in weakly coupled spin dimers and chains material \CuFeGeO}
\author{T. Masuda}
\email[]{tmasuda@yokohama-cu.ac.jp}
\altaffiliation{ Present address: International Graduate School of Arts and Sciences,
Yokohama City University, 22-2, Seto, Kanazawa-ku,
Yokohama city, Kanagawa, 236-0027, Japan}
\affiliation{Condensed Matter Sciences Division, Oak Ridge National
Laboratory, Oak Ridge, TN 37831-6393, USA}

\author{A. Zheludev}
\affiliation{Condensed Matter Sciences Division, Oak Ridge National
Laboratory, Oak Ridge, TN 37831-6393, USA}

\author{B. Sales}
\affiliation{Condensed Matter Science Division, Oak Ridge National
Laboratory, Oak Ridge, TN 37831-6393, USA}

\author{S. Imai}
\altaffiliation{ Present address: Semiconductor R \& D Center,
Yokohama R \& D Laboratories,
 The Furukawa Electric Co., Ltd. }
\affiliation{Department of Advanced Materials Science, The University
of Tokyo, 5-1-5, Kashiwa-no-ha, Kashiwa 277-8581, Japan}

\author{K. Uchinokura}
\altaffiliation{ Present address: The Institute of Physical and
Chemical Research (RIKEN), Wako, Saitama 351-0198, Japan}
\affiliation{Department of Advanced Materials Science, The University
of Tokyo, 5-1-5, Kashiwa-no-ha, Kashiwa 277-8581, Japan}

\author{S. Park}
\altaffiliation{ Present address: HANARO Center, Korea Atomic Energy Research 
Institute, Daejeon, South Korea}
\affiliation{NIST Center for Neutron Research, National Institute
of Standards and Technology, Gaithersburg, Maryland 20899, USA}

\date{\today}

\begin{abstract}
Magnetic excitations in a weakly coupled spin dimers and chains
compound \CuFeGeO\ are measured by inelastic neutron scattering.
Both structure factors and dispersion of low energy excitations up
to 10 meV energy transfer are well described by a semiclassical
spin wave theory involving interacting Fe$^{3+}$ ($S = 5/2$)
chains. Additional dispersionless excitations are observed at
higher energies, at $\hbar \omega = 24$ meV, and associated with
singlet-triplet transitions within Cu$^{2+}$-dimers. Both types of
excitations can be understood by treating weak interactions
between the Cu$^{2+}$ and Fe$^{3+}$ subsystems at the level of the
Mean Field/ Random Phase Approximation. However, this simple model
fails to account for the measured temperature dependence of the
24~meV mode.
\end{abstract}

\pacs{75.10.Jm, 75.25.+z, 75.50.Ee}

\maketitle

\section{introduction}
The ground states of low-dimensional  magnets are strongly
affected by quantum spin fluctuation. In so-called quantum spin
liquids spin correlation remains short-range even at zero
temperature, and the excitation spectrum is gapped. This disorder
is relatively robust and such systems resist long-range ordering
even in the presence of residual 3D interactions or anisotropy. In
contrast, gapless low-dimensional magnets are very sensitive to
external perturbations that can easily drive them towards
long-range ordering. New physics is found in bicomponent systems
that combine these two distinct types of low-dimensional spin
networks. An example is $R_2$BaNiO$_5$ materials where Haldane
spin chains weakly interact with magnetic rare earth ions.
~\cite{Zheludev96a,Zheludev98a,Alvarez02} More recently, we
reported the discovery and study of a novel quantum ferrimagnet
\CuFeGeO .\cite{Masuda03a,Masuda04a} We showed that this compound
can be viewed as a system of antiferromagnetic (AF) Cu-dimers that
weakly interact with almost classical Fe$^{3+}$ chains. This weak
coupling leads to a rather unusual {\it cooperative} ordering
phase transition at low temperatures.

The crystal structure of \CuFeGeO\ is monoclinic $P2_1/m$ with $a$
= 12.101 \AA , $b$ = 8.497 \AA, $c$ = 4.869 \AA, and $\beta$ =
96.131$^{\circ}$.\cite{Masuda04a} The arrangement of magnetic ions
and likely exchange pathways is shown in Fig.~\ref{fig1} (a).
Fe$^{3+}$ ions form crankshaft-shaped chains that run in the $b$
direction. These chains are separated by GeO$_4$ tetrahedra in $c$
direction. The Cu$^{2+}$ dimers are located in-between Fe$^{3+}$
chains along the $a$ direction. Simultaneous cooperative
long-range magnetic ordering of the two magnetic subsystems occurs
at 40 K.\cite{Masuda04a} The magnetic structure is roughly
collinear, with spins lying in the crystallographic $a$ - $c$
plane. In addition, there are small out-of-plane spin components
(canting). The saturation magnetic moment on Cu$^{2+}$ ions is
anomalously small, being suppressed by residual quantum
fluctuations in the dimers: $m_{\rm Cu}~=~0.38(4) \mu_{\rm B}$.
This suggests that the coupling $J_{\rm Cu-Fe}$ between the
Cu-dimers to the Fe-subsystem is weak compared to intra-dimer AF
interactions $J_{\rm Cu}$. The pairs of Cu-spins remain in a
spin-singlet state that is only partially polarized by
interactions with the long-range order in the Fe-subsystem. The
data collected in preliminary inelastic neutron scattering
experiments for energy transfers up to 10 meV could be well
explained by fluctuations of Fe$^{3+}$ spins alone. The Fe$^{3+}$ spins 
form weakly-coupled $S=5/2$ chains, the corresponding effective
exchange constants being $J_{\rm Fe}~=~1.60(2)$ meV, $J'_{\rm
Fe}~=~0.12(1)$ meV, as shown in Fig.~\ref{fig1}. These values are
consistent with rough estimates of exchange constants based on
magnetic susceptibility data: $J_{\rm Fe}~=~1.7$ meV, $J_{\rm
Cu}~=~25$ meV. To date, no excitations associated with the
Cu-dimers could be identified.

While the very nature of the crankshaft-shaped chains implies some
alternation of bond strength, we find no evidence thereof in the measured
dispersion curves. In fact, they are well reproduced by a model
involving uniform magnetic Fe-chains with weak interchain interactions along the 
$c$ direction. In the remainder of this work we shall therefore disregard the
bond-alternation and assume the Fe-chains to be magnetically uniform. 

In the present paper we report a more detailed inelastic neutron
scattering study of \CuFeGeO.  Our main result is an observation
of a complete separation of energy scales for Cu$^{2+}$- and
Fe$^{3+}$-centered magnetic excitations. An analysis of the
intensity pattern for low-energy spin waves allows us to
unambiguously associate them with the dynamics of $S=5/2$ chains.
In addition, separate narrow-band excitation originating from Cu
subsystem is observed at  higher energy transfers. The layout of
the paper will be as follows. In section \ref{experimental} we
will describe the characteristics of our samples  and experimental
setups. Section \ref{results} describes the results for low-energy
excitations in single-crystal samples, as well as measurements of
the much weaker high energy excitations in a large-volume powder
sample. In Section IV we shall interpret the observed separation
of energy scales by a Mean Field- Random Phase Approximation
(MF-RPA) treatment of interactions between Cu$^{2+}$- and
Fe$^{3+}$ spins. Discussion and conclusion will be drawn in 
Section V and VI, respectively.
\section{experimental}
\label{experimental}

\begin{figure}
\begin{center}
\includegraphics[width=8.5cm]{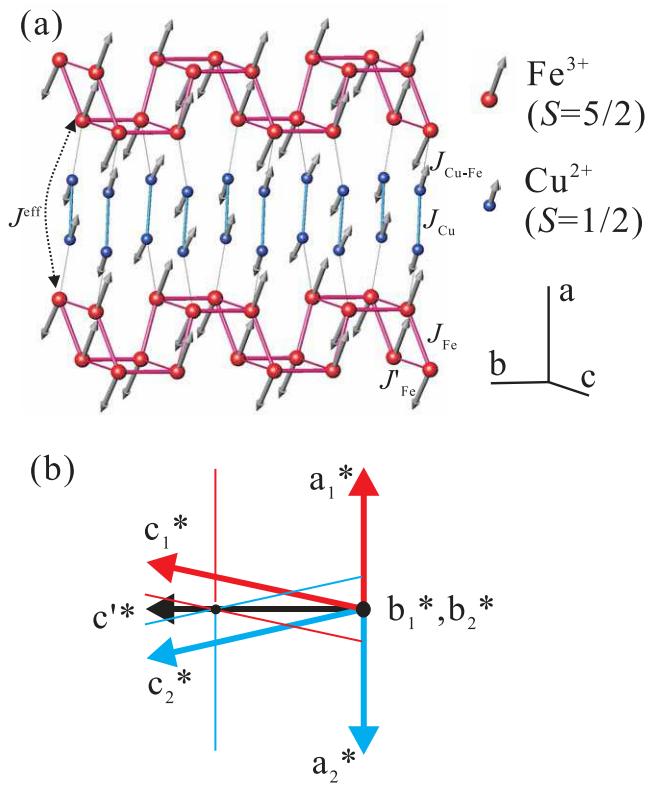}
\end{center}
\caption{(a) A schematic view of the crystal structure of
Cu$_2$Fe$_2$Ge$_4$O$_{13}$. Only magnetic ions are shown. Arrows
symbolize the spin structure in the magnetically ordered phase.
Possible exchange pathways are also shown. (b) Relative
orientations of reciprocal-lattice vectors for the two domains in
bulk crystalline samples with twinning.} \label{fig1}
\end{figure}

High quality single crystals with the dimension of 
$3\times 4 \times 35$ mm$^3$
were grown by the floating zone method. All
samples were found to be  twinned. The two twins share a common
$(b,c)$ plane, the monoclinic structure allowing two independent
orientations of the $a$ axis. In reciprocal space the domains
share a common $(a^{*},b^{*})$ plane, but have distinct $c^\ast$
axes, as illustrated in Fig.~\ref{fig1} (b). Two separate single
crystals were co-aligned to obtain  a larger sample of cumulative
mosaic spread 0.44$^\circ$.  For single crystal inelastic neutron
scattering experiments we exploited four different setups. Setup
I employed the SPINS cold neutron spectrometer at the NIST Center
for Neutron Research (NCNR). The scattering plane was defined by
the $b^{*}$ and the bisector $c^{'*}$ of the $c^*$ axes in the two
crystallographic domains, as shown in Fig.~\ref{fig1} (b). In this
geometry the scattering planes are {\it the same} in both domains.
It is convenient to define $c^{'*} \equiv c^*\cos (\beta
-90^{\circ})$. The momentum transfer in the scattering plane of
the spectrometer is then indexed by Miller indexes $h'$, $k'$ and
$l'$ of a  fictitious orthorhombic structure. For the two domains
$h = \pm (c^{'*}/a^{*})l'\tan (\beta-90^{\circ}), k = k'$ and
$l=l'$. Since $\beta$ is close to 90$^{\circ}$, $h$ is almost zero
for most measurements using Setup 1, where $l$ is small. In Setup
 2 the scattering plane was $(a^{*},b^{*})$, which is also common for
the  two domain types. In both setups we used $\mathrm{(guide)} -
80' -  80' -  \mathrm{(open)}$ collimation with a Be filter
positioned after the sample and a fixed final energy $E_{\rm f} =
5$~meV or 3~meV. The data were collected at $T$ = 1.4 K using a
standard He-flow cryostat. Setup 3 was used for wide surveys in
reciprocal space and employed the HB1 thermal neutron spectrometer
at the High Flux Isotope Reactor at ORNL. The scattering plane was
$(a^{*},b^{*}$) , as in Setup 2. The collimation was $48' - 40' -
40' - 240'$. Neutrons
with $E_f$ = 13.5 meV were used in conjunction with  a Pyrolytic
graphite (PG) filter positioned after the sample. The experiments
in Setup 3 were performed at $T = 6.4$~K maintained by a
closed-cycle He refrigerator.

Intensity was the main limiting factor for studies of
higher-energy magnetic excitations. To maximize sample volume the
measurements were performed on a 50~g polycrystalline \CuFeGeO\ 
powder that was prepared by the solid state reaction method. This
experiment was performed on the HB3 3-axis spectrometer at HFIR
with $48' - 40' - 60' - 120'$ collimations and  a PG filter after
the sample (Setup 4). The final neutron energy was fixed at
$E_f$ = 14.7 meV. A closed-cycle refrigerator was used to achieve
low temperatures.

\section{Experimental Result}
\label{results}
\subsection{Low energies}
\begin{figure}
\begin{center}
\includegraphics[width=8.5cm]{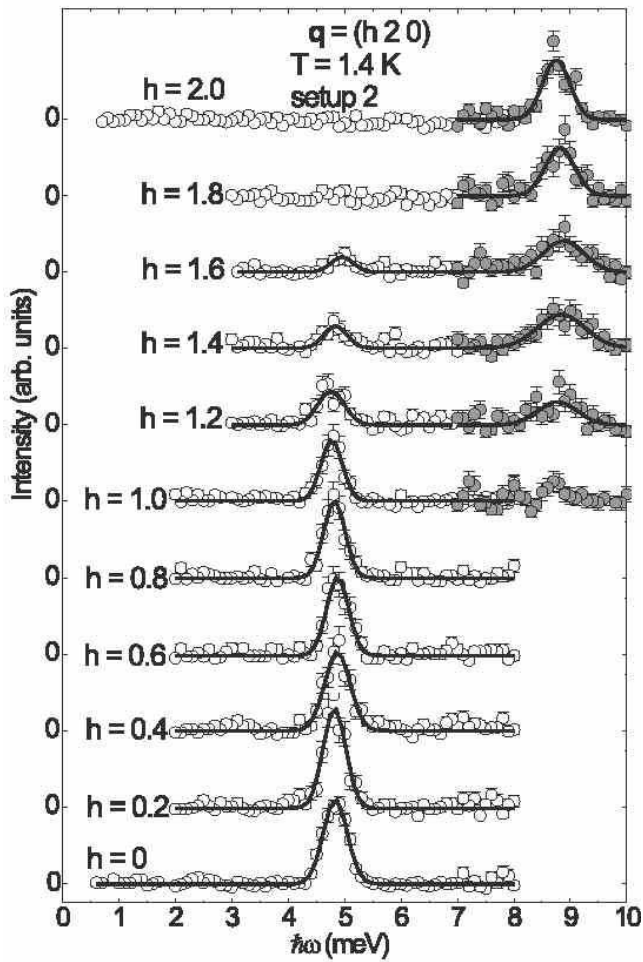}
\end{center}
\caption{ Energy
scans collected in \CuFeGeO\ for momentum transfers ${\mathbf q} =
(0~ k~ 0.5)$. White and grey circles correspond to measurements
with $E_{\rm f}$ = 5 and 3 meV, respectively. Solid lines are
Gaussian fits. \label{fig2}}
\end{figure}
In this section we concentrate on the low energy excitations at
energy transfers up to 10~meV. As was explained in our preliminary
report\cite{Masuda04a} and will be discussed in detail below, this
part of the spectrum can be associated with conventional spin
waves from the Fe-subsystem. The Cu-dimers only provide an
effective Fe-Fe interaction, but contribute nothing to the
dynamics at low energies.

\begin{figure}
\begin{center}
\includegraphics[width=8.5cm]{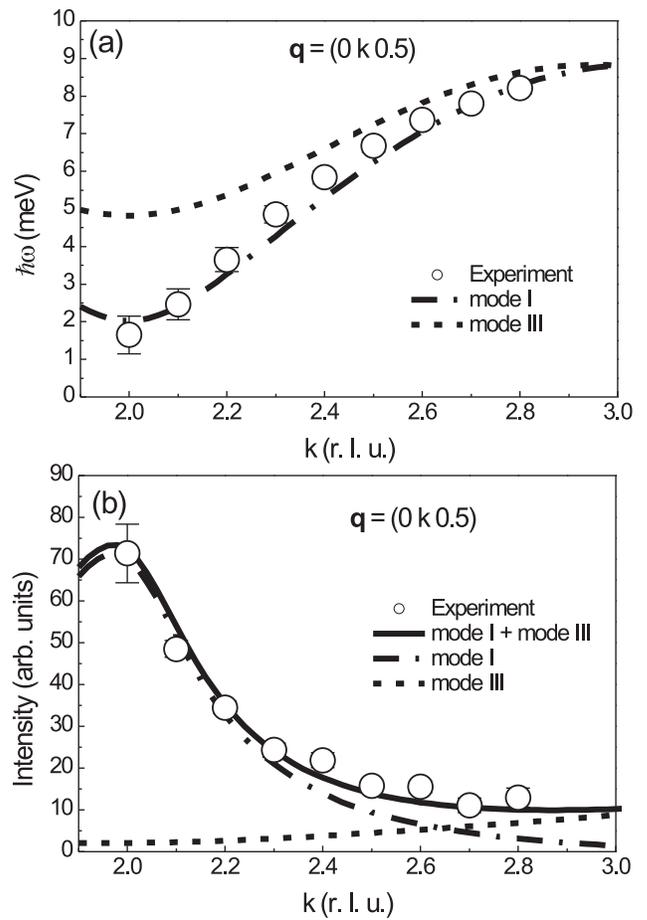}
\end{center}
\caption{\label{fig3} Measured energies (a)  and integrated
intensities (b) of magnetic excitations as a function of momentum
transfer along the $b^{*}$ direction (symbols). Lines are as
described in the text. }
\end{figure}

Typical energy scans at ${\mathbf q} = (0~k~0.5)$ measured using
setup 1 are shown in Fig.~\ref{fig2}. White and grey circles
correspond to data collected with $E_{\rm f}$ = 5 meV and 3 meV
respectively. Well defined resolution-limited peaks are observed
in the entire Brillouin zone. Solid lines are Gaussian fits to the
data after a subtraction of a linear background. The excitation
energy is a minimum at the magnetic zone center at ${\mathbf q} =
(0~2.0~0.5)$. A small gap of about 1.6~meV is observed at this
wave vector. The apparent shoulder structure can be attributed to
a splitting of the spin wave branch into two components with
somewhat different gap energies. The gaps are most likely
anisotropy-related.  The zone-boundary energy at
$\mathbf{q}=(0~3.0~0.5)$ is about 9~meV. The dispersion relation
was obtained by Gaussian fits to individual scans and is shown in
Fig.~\ref{fig3} (a). The $k$-dependence of the measured
energy-integrated peak intensity is plotted in symbols in
Fig.~\ref{fig3} (b). The intensity scales roughly as $1/\omega$.

\begin{figure}
\begin{center}
\includegraphics[width=8.5cm]{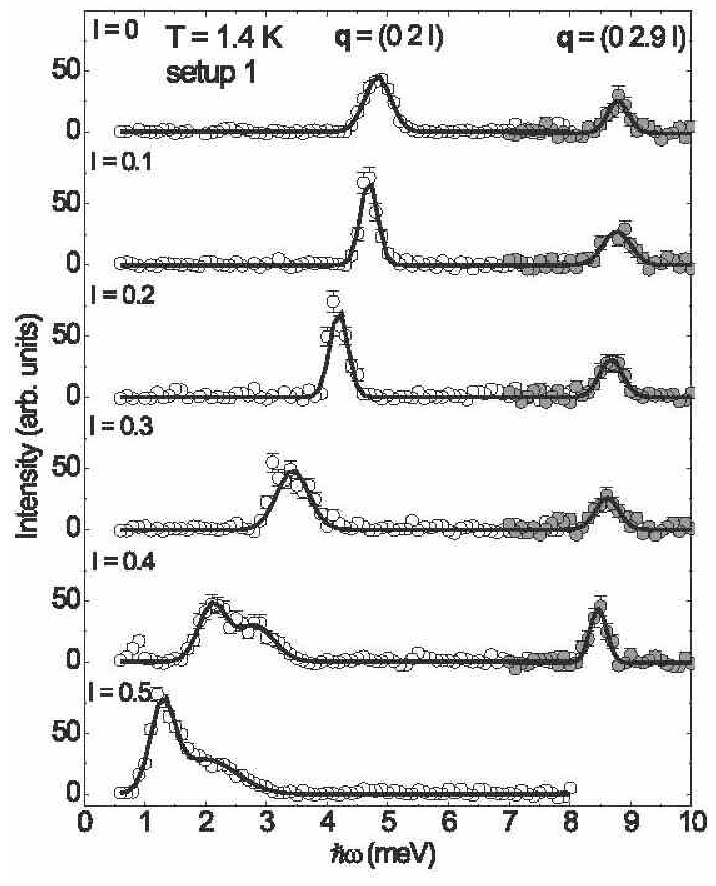}
\end{center}
\caption{\label{fig4} Energy scans collected in \CuFeGeO\ for
several momentum transfers along the $c^{*}$ direction. Symbols
and lines are as in Fig.~\protect\ref{fig2}.}
\end{figure}

\begin{figure}
\begin{center}
\includegraphics[width=8.5cm]{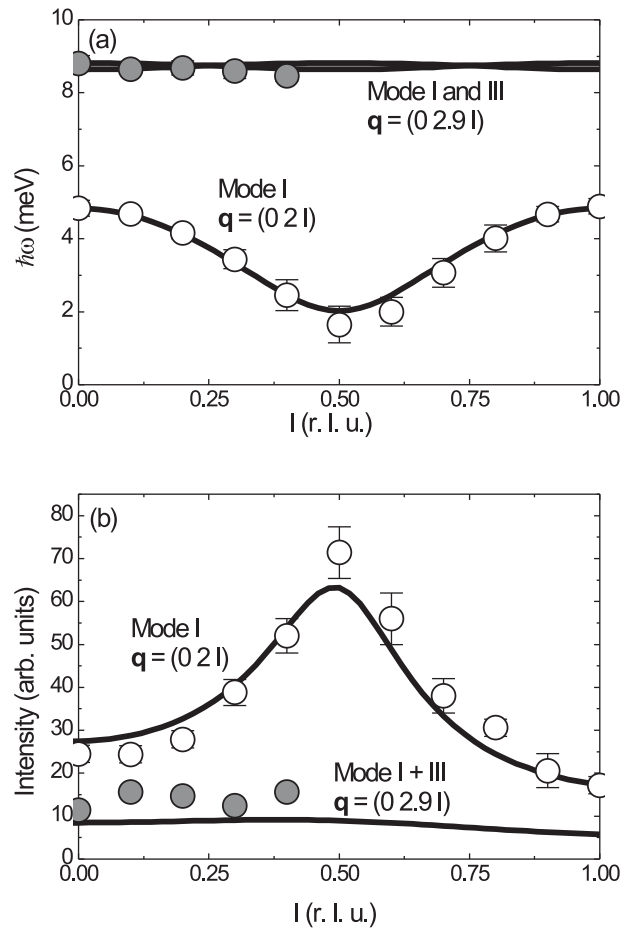}
\end{center}
\caption{\label{fig5} Measured energies (a)  and integrated
intensities (b) of magnetic excitations as a function of momentum
transfer along the $c^{*}$ direction.
 The solid and open circles correspond to the higher- and
lower-energy branches, respectively. Lines are as described in the
text.}
\end{figure}

Energy scans at ${\mathbf q} = (0~2.0~l)$ and $(0~2.9~l)$ are
shown in Fig.~\ref{fig4}. Constraints on experimental geometry
prevented us from reaching the more symmetric $(0~3.0~l)$
reciprocal-space rods at higher energy transfers.
These two sets of data reveal two distinct excitation branches,
that are strongest near even  and odd $k$-values, respectively.
The former branch is dispersive along the $c$ axis, while the
latter one is almost flat. Dispersion relations and the integrated
intensity plots for both modes were obtained using Gaussian fits
and are shown in Fig.~\ref{fig5} (a) and (b).
Even for the more dispersive lower-energy branch the booundary energy
along the $c^{'*}$ direction is only about 5 meV: about half of
that in the $b^{*}$ direction.

\begin{figure}
\begin{center}
\includegraphics[width=8.5cm]{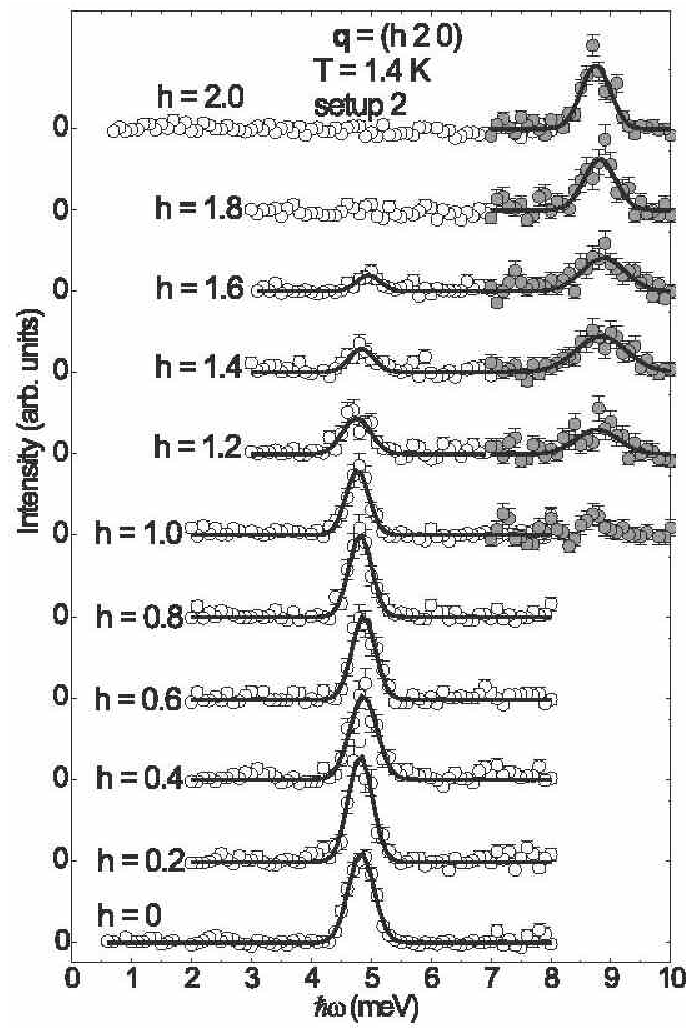}
\end{center}
\caption{\label{fig6}Energy scans measured in \CuFeGeO\ for
momentum transfers ${\mathbf q} = (h~2.0~0)$. Symbols and lines
are as in Fig.~\protect\ref{fig2}.}
\end{figure}

No dispersion of low-energy excitations could be detected along
the $a^{*}$ direction. This is illustrated by the energy scans
collected on the $(h~2.0~0)$ reciprocal-space rod using Setup 2
(Fig.~\ref{fig6}).
\begin{figure}
\begin{center}
\includegraphics[width=8.5cm]{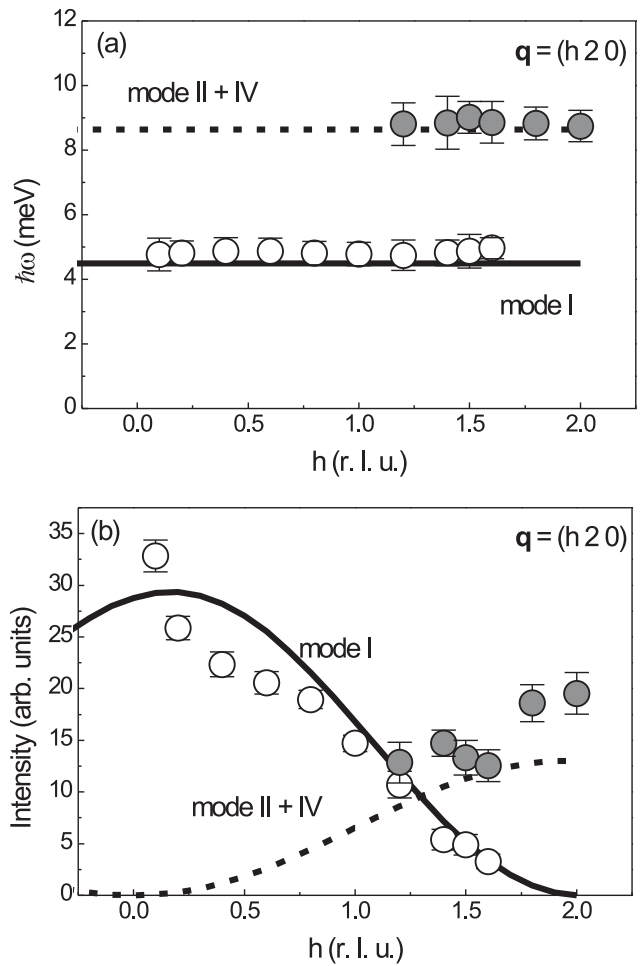}
\end{center}
\caption{\label{fig7} (a) Measured energies (a)  and integrated
intensities (b) of magnetic excitations as a function of momentum
transfer along the $a^{*}$ direction. Symbols and lines are as in
Fig.~\protect\ref{fig5}. }
\end{figure}
Two peaks are observed at 5 and 9 meV, respectively. The
excitation energies are plotted in Fig.~\ref{fig7} (a).
Interestingly, the peak intensity is strongly dependent on $h$, as
shown in Fig.~\ref{fig7} (b).

From the data presented above one can immediately conclude that
the most relevant magnetic interactions are within the Fe$^{3+}$
layers, parallel to the $(b,c)$ plane. Inter-layer coupling along
the $a$-direction is considerably weaker. Based on structural
considerations, it must involve the Cu$^{2+}$ spins.

\subsection{Higher-energy excitations}
The model that we previously proposed for \FeCuGeO\ 
(Ref. \onlinecite{Masuda04a}) implies a separation of energy scales of spin
wave like excitations on the Fe subsystem and triplet excitations
of the Cu dimers. From magnetic susceptibility measurements we
have estimated the intra-dimer exchange constant to be around
25~meV. Due to small sample size, in single crystal experiments we
failed to observe any clear magnetic inelastic features in this
energy transfer range. To search for the Cu-triplet mode we
performed additional measurements on a large-size powder sample.

\begin{figure}
\begin{center}
\includegraphics[width=8.5cm]{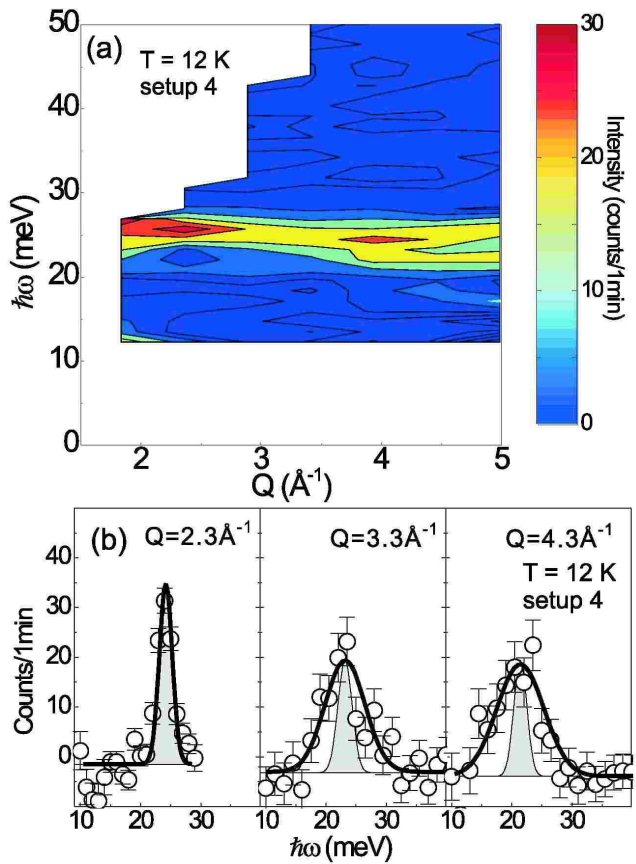}
\end{center}
\caption{(a) False-color plot of inelastic intensity measured in
\CuFeGeO\ powder samples as a function of energy and momentum
transfer. (b) Typical energy scans measured at $T=12$~K for
different momentum transfers. Heavy solid lines are Gaussian fits.
The shaded Gaussians represent experimental energy resolution.
}
\label{fig10}
\end{figure}

The powder data collected at $T=12$~K are summarized in the false
color plot in Fig.~\ref{fig10} (a).  The data were obtained by
combining 17 separate constant-$q$ scans. For each such scan a
linear background was subtracted from the measured intensity.
Figure~\ref{fig10} (a) clearly shows a narrow excitation band at
$\hbar \omega \sim 24$~meV. The measured intensity of the 24~meV
peak decreases with the increase of $q$, as expected for magnetic
scattering. The observed energy width at $q = 2.3$ \AA$^{-1}$ is
somewhat larger than experimental resolution, but still small
compared to the central energy (Fig.~\ref{fig10} (b)). Such energy
dependence in powder samples typically indicates a narrow
dispersion bandwidth (see, for example,
Refs.~\onlinecite{Zheludev96b,Masuda03b}). The observed spectrum
is consistent with the excitations originating from the structural
Cu-dimers, indicated by the thick solid bonds in Fig.~\ref{fig1}
(a). However, the momentum transfer range covered in the
experiment, especially at $|q|\rightarrow 0$ is insufficient for a
more detailed analysis of the structure factor. In particular, the
size of the dimers can not be independently extracted from the
experimental data.

To obtain additional information on the 24~meV excitation we
studied its temperature dependence at $q$ = 2.3 \AA $^{-1}$. The
main challenge was dealing with a large background that originates
from (i) temperature-independent scattering, including spurious
scattering from an  ``accidental'' Bragg powder line at 24.8~meV
and (ii) temperature-dependent phonon scattering. These two
contributions can be removed from the data if one assumes that the
useful magnetic signal is relatively weak and/ or
temperature-independent above $T=200$~K. This would indeed be true
if the magnetic scattering originated from effectively isolated
Cu-dimers with an intradimer exchange constant of 24~meV. The
phonon contribution, which is assumed to scale with the Bose
factor, is thus estimated from comparing scans at $T=300$~K and
$T=200$~K. It is then appropriately scaled and subtracted from all
scans, leaving only the true magnetic signal and the
$T$-independent background. The two can not, in principle, be
reliably separated. However, we can follow the {\it change} in
magnetic signal from $T=200$~K by using the phonon-subtracted
$T=200$~K scan as ``background''.

\begin{figure}
\begin{center}
\includegraphics[width=8.5cm]{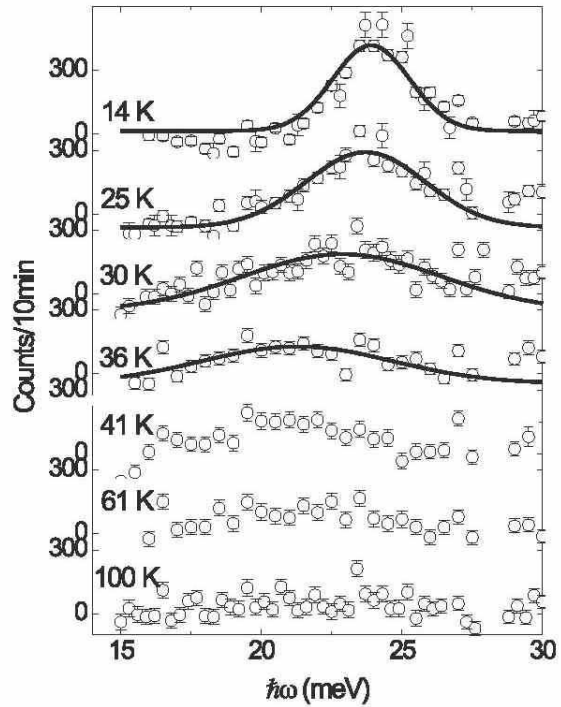}
\end{center}
\caption{Typical energy scans at measured in \CuFeGeO\ powder at
$q$ = 2.3 \AA $^{-1}$ for different temperatures. Solid lines are
Gaussian fits. The background has been subtracted, as explained in
the text.} \label{fig12}
\end{figure}

The result of this elaborate background subtraction is plotted in
Fig.~\ref{fig12}. The well-defined peak seen at low temperature
broadens progressively with increasing $T$ and practically
disappears at $T \gtrsim 40$~K. In this range there also seems to
be a downward shift in the peak's central energy.


\section{Analysis}

\subsection{Separation of energy scales}
To understand the dynamics of the coupled Fe$^{3+}$ and Cu$^{2+}$
subsystems we shall make the central assumption that the exchange
constant $J_\mathrm{Cu}$ that binds pairs of Cu$^{2+}$ spins into
AF dimers is the largest energy scale in the system. Under these
circumstances the degrees of freedom associated with Cu-spins can
be effectively integrated out at low energies. Indeed, the
isolated Cu$^{2+}$-subsystem has a spin singlet ground state and a
large energy gap $\Delta=J_\mathrm{Cu}$. At $\hbar \omega \ll
\Delta$ it lacks any intrinsic dynamics, i.e., its dynamic
susceptibility is purely real and almost energy-independent. In
the spirit of RPA, from the point of view of Fe$^{3+}$ spins, the
Cu-dimers merely act as a polarizable medium that can transfer
magnetic interactions between the Fe-layers. The staggered spin
susceptibility of each $S=1/2$ dimer being $2/(J_\mathrm{Cu})$,
this effective coupling, labeled as $J_\mathrm{eff}$ in
Fig.~\ref{fig1} is given by:
\begin{equation}
J^\mathrm{eff}=J_\mathrm{Cu-Fe}^2/(2 J_\mathrm{Cu}).
\end{equation}
Thus, to a good approximation, the low-energy spin dynamics of
\FeCuGeO\ is simply that of the Fe$^{3+}$-subsystem with an
additional exchange coupling. From the experiment, where no dispersion
of spin waves could be observed along the $a$-axis, the effective
exchange constant must be rather small. Nevertheless, as explained
in our previous paper,\cite{Masuda04a} it is absolutely crucial in
completing a 3-dimensional spin network and allowing long-range
magnetic ordering at a non-zero temperature.

At high energy transfers, comparable to the Cu-dimer gap, it is
the Fe$^{3+}$ degrees of freedom that can be effectively
integrated out. At $T<T_{\mathrm{N}}$ their effect is reduced to
producing a {\it static} staggered exchange spin field that acts
on the Cu$^{2+}$ spins and is proportional to the ordered
Fe$^{3+}$ moment:
 \begin{eqnarray}
 h^\mathrm{Cu} &=& \langle S_\mathrm{Fe} \rangle J_{\rm Cu-Fe}.
 \label{JCuFe}
 \end{eqnarray}
In our approximation at high energy transfers \CuFeGeO\ behaves as
a collection of (possibly interacting) Cu-dimers in an effective
staggered field.\cite{Masuda04a}

\subsection{Spin waves and dynamic structure factor for low energies}
\begin{figure}
\begin{center}
\includegraphics[width=8.5cm]{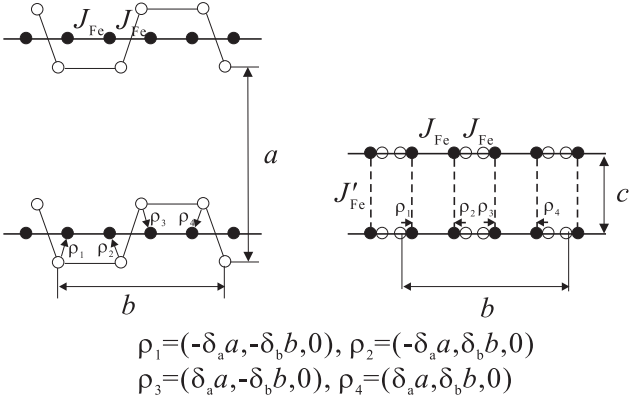}
\end{center}
\caption{Spatial arrangement of Fe$^{3+}$-sites in the structure
of \CuFeGeO (open circles) and an equivalent Bravais lattice of
spins (solid circles).}
\label{fig9b}
\end{figure}
As explained above, the low-energy spectrum of \CuFeGeO\ can be
understood by considering the Fe-subsystem in isolation, and even
$J_\mathrm{eff}$ can be ignored for its smallness. For the
magnetically ordered state the dynamic structure factor of such a
system can be calculated using conventional spin wave theory
(SWT). For simplicity we shall assume a collinear magnetic
structure, ignoring the small canting of the Fe$^{3+}$ spins out
of the $(a,c)$ plane.

Even though there are four Fe ions in each unit cell of \CuFeGeO,
the topology of exchange interactions defined by $J_\mathrm{Fe}$
and $J'_\mathrm{Fe}$ in Fig.~\ref{fig1} is that of a regular
Bravais lattice. This equivalency is illustrated in
Fig.~\ref{fig9b}. The {\it exact} relation between the actual
structure factor $S(\mathbf{q},\omega)$ and
$S_0(\mathbf{q},\omega)$ for the equivalent lattice is readily
obtained:
 \begin{subequations}
 \begin{widetext}
 \begin{eqnarray}
 S(\mathbf{q},\omega)&=&S_0(\mathbf{q},\omega)\cos^{2}2\pi\delta_{a}h\cos^{2}2\pi\delta_{b}k
 +\frac{1}{2}S_0(\mathbf{q}+(0,1,0),\omega)(1+\sin^22\pi\delta_{a}h\sin^{2}2\pi\delta_{b}k)
 \nonumber\\
 &+&
 S_0(\mathbf{q}+(0,2,0),\omega)\cos^{2}2\pi\delta_{a}h\sin^{2}2\pi\delta_{b}k
 +\frac{1}{2}S_0(\mathbf{q}+(0,3,0),\omega)(1+\sin^22\pi\delta_{a}h\sin^{2}2\pi\delta_{b}k),\label{zigzagterms}
 \end{eqnarray}
 \end{widetext}
 \begin{eqnarray}
 \delta_{a}=0.1239,
 \delta_{b}=0.0629.
 \end{eqnarray}
 \label{zigzag}
 \end{subequations}
Here $\delta_a$ and $\delta_b$ are the displacements of Fe$^{3+}$
ions from high-symmetry positions, as shown in Fig.~\ref{fig1}
(b).

The SWT dynamic structure factor for a collinear  antiferromagnet
on a Bravais lattice is well known:\cite{Lovesey}
\begin{subequations}
 \begin{eqnarray}
 S_0(\mathbf{q},\omega)&=&(u_{\mathbf{q}}+v_{\mathbf{q}})^2
 \delta(\hbar \omega _{\mathbf{q}}-\hbar \omega) \label{uniform}\\
 u_{\mathbf{q}}^2&=&\frac{S(\hbar \omega _{\mathbf{q}}+2Sj(0))}
 {\hbar \omega _{\mathbf{q}}},\\
 u_{\mathbf{q}}v_{\mathbf{q}}&=&-\frac{2S^2j(\mathbf{q})}{\hbar \omega
 _{\mathbf{q}}},\\
 \hbar \omega _q&=&S\sqrt{j(0)^2-j(\mathbf{q})^2 + \Delta ^2}.\label{dispersion}
 \end{eqnarray}
 \end{subequations}
Here $j(\mathbf{q})$ is the Fourier transform of exchange
interactions, and $\Delta$  empirically accounts for the
anisotropy gap. In our particular case of the Fe$^{3+}$ subsystem
in \CuFeGeO\ we have:
 \begin{equation}
 j(\mathbf{q})=2(J_{\rm Fe} \cos \frac{\pi}{2} k + J'_{\rm Fe} \cos 2\pi l), \label{jq}
 \end{equation}
The cross section for inelastic neutron scattering from spin waves
is given by:
\begin{equation}
 \frac{d^2\sigma}{d\Omega dE}
 \propto |F(\mathbf{q})|^2  \left[1+\left(\frac{q_y}{q}\right)^2\right]
 \langle n_{\mathbf{q}}+1 \rangle S(\mathbf{q},\omega),\label{cross}
\end{equation}
In this formula $F(\mathbf{q})$ is the magnetic form factor for
Fe$^{3+}$, $\langle n_{\mathbf{q}} \rangle$ is the Bose factor and
$q_z$ is the projection of the scattering vector onto the
direction of ordered Fe$^{3+}$ moments.

\subsection{Fits to data}
 The model cross section given by Eq.~(\ref{cross}) can
accurately reproduce the observed low-energy spectra in \CuFeGeO.
Due to the presence of four terms in Eq.~(\ref{zigzag}), there are
four distinct spin wave branches, that we shall denote as modes I
through IV, correspondingly. The dispersion relation given by
Eq.~(\ref{dispersion}) was fit to the experimental data shown  in
Figs.~\ref{fig3} (a) and \ref{fig5} (a) using a least-squares
algorithm. A good fit is obtained with $J_{\rm Fe} = 1.60$~meV,
$J'_{\rm Fe} = 0.12$~meV, and $\Delta = 2.02$~meV. The result is
shown in  lines in figures \ref{fig3}a and \ref{fig5}a. With these
parameters our model also agrees well with the measured dispersion
(or, rather, absence thereof) along the $a$ axis, as shown in
\ref{fig7}a.

\begin{figure}
\begin{center}
\includegraphics[width=8.5cm]{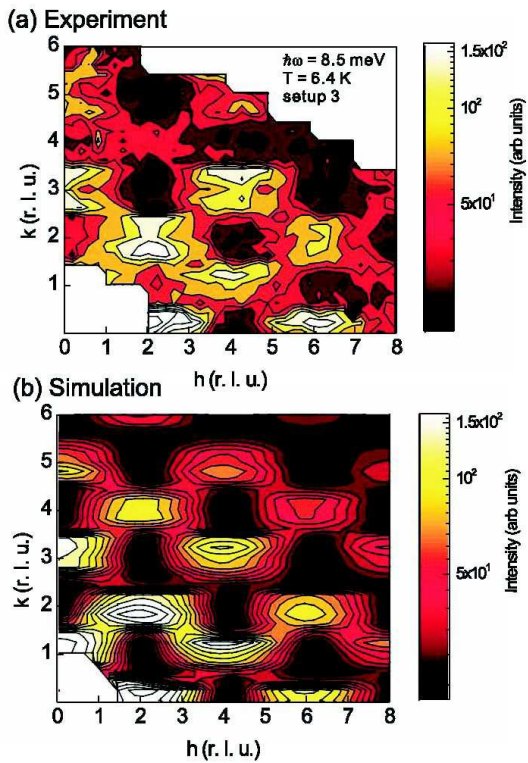}
\end{center}
\caption{(a) Constant energy scans at $\hbar \omega =$ 8.5 meV
in wide ${\mathbf q} = (h~k~0)$ range.
 (b) Simulation based of our model cross section convoluted
with the experimental resolution function. } \label{fig8}
\end{figure}

What is important, is that not only the energies, but also the
{\it intensities} of the observed excitations are well reproduced
by our model. Calculated intensities for each mode or combined
intensities of a couple of modes in cases where experimental
energy resolution is insufficient to resolve individual branches,
are shown in lines in Figs.~\ref{fig3}b, \ref{fig5}b and
\ref{fig7}b. This quantitative agreement between the measured and
observed structure factors confirms that the scattering is indeed
due to Fe$^{3+}$ spins. An excellent illustration of this was
obtained by mapping out the scattering intensity in a wide
$\mathbf{q}$-range using Setup 4, as shown in Fig.~\ref{fig8}
(a). These data correspond to a fixed energy transfer $\hbar
\omega = 8.5$~meV, and have a characteristic checkerboard pattern.
Our spin wave model with the parameters quoted above reproduces
this behavior very well, as shown in Fig.~\ref{fig8} (b).  In this
calculation the model cross section was numerically convoluted
with the known experimental resolution function. The apparent
periodicity along $k$ is due to a steep dispersion that takes the
excitations in and out of the probed energy range. However, the
periodicity along $h$ is related to the trigonometric coefficients
in Eq.~(\ref{zigzagterms}). These, in turn, are determined by the
geometry of the crankshaft-shaped Fe$^{3+}$ chains in \CuFeGeO.

\section{Discussion}
As demonstrated above, the low-energy spin dynamics of \CuFeGeO\
is well described by an effective spin wave theory for the
Fe$^{3+}$ spin chains. Based on the available data it is impossible to
unambiguously associate the observed 24~meV mode with the
Cu$^{2+}$ dimers. However, much confidence in this assumption can
be drawn from a recent study of \CuScGeO .\cite{Masuda05unpublished} 
In this {\it isostructural}
compound\cite{Redhammer04a} only the Cu$^{2+}$ ions are magnetic.
Indeed the magnetic susceptibility 
showed $S=1/2$-dimers behavior with similar energy to 
\CuFeGeO .\cite{Masuda05unpublished}

We can now estimate all the relevant exchange interactions in the
system. Using Eq.~(\ref{JCuFe}) in combination with the staggered 
susceptibility of an isolated antiferromagnetic dimer, 
from the known saturation moments 
$\langle S_{\rm Cu} \rangle~=~0.18(8)$ and $\langle S_{\rm
Fe} \rangle~=~1.77(4)$ for Cu$^{2+}$ and Fe$^{3+}$,
respectively, \cite{Masuda04a} we get $J_{\rm Cu-Fe}~=~2.54(3)$~meV.
The effective coupling is then 
$J^{\rm eff}~=~0.13(4)$~meV. The results for all exchange parameters
are summarized in Table~\ref{parameters}. 

\begin{table*}
 \caption{Exchange parameters for \CuFeGeO.}
 \label{parameters} 
 \begin{ruledtabular}
 \begin{tabular}{l l l l l l}
 $J_{\rm Fe}$ & $J'_{\rm Fe}$ & $J^{\rm eff}$ & $J_{\rm Cu-Fe}$ & $J_{\rm Cu}$ \\
 \hline
 1.60(2)~meV & 0.12(1)~meV & 0.13(4)~meV& 2.54(3)~meV &24.(2)~meV
 \end{tabular}
 \end{ruledtabular}
 \end{table*}

Our model for \CuFeGeO\ is qualitatively consistent with the
observed slight increase of the energy of the Cu-dimer mode with
decreasing temperature. Indeed, the gap energy of isolated dimers,
as that of other gapped systems, is known to increase with the
application of a staggered field.\cite{Jolicoeur94a,Ma95a} In
\CuFeGeO\ this field is generated by the ordered moment on the
Fe$^{3+}$ sites, and at $T<T_\mathrm{N}$ increases proportionately
to $\langle S_\mathrm{Fe} \rangle$. Beyond that, the observed
$T$-dependence of the 24~meV mode is  different from that for
isolated dimers. In the latter, the intensity would remain almost
constant below $T=40$~K. Moreover, the peak would remain sharp at
all temperatures. The discrepancy may be related to intrinsic
limitations of the MF/RPA approach, and merits further
investigation. In particular, it is tempting to somehow associate
the observed emergence and sharpening of the 24~meV inelastic peak
with the onset of long-range order.

\section{Conclusion}
 Our data bring solid quantitative support to the
concept of separation of energy scales in the mixed-spin quantum
antiferromagnet \CuFeGeO. Using a simple MF/RPA approach we are
able to determine all the relevant exchange interactions. However,
certain features of the temperature dependence of spin excitations
require further theoretical and experimental study.

\begin{acknowledgments}
Work at ORNL was carried out under Contracts No.
DE-AC05-00OR22725, US Department of Energy. Experiments at NIST
were supported by the NSF through DMR-0086210 and DMR-9986442.
\end{acknowledgments}


\begin{thebibliography}{10}
\expandafter\ifx\csname natexlab\endcsname\relax\def\natexlab#1{#1}\fi
\expandafter\ifx\csname bibnamefont\endcsname\relax
  \def\bibnamefont#1{#1}\fi
\expandafter\ifx\csname bibfnamefont\endcsname\relax
  \def\bibfnamefont#1{#1}\fi
\expandafter\ifx\csname citenamefont\endcsname\relax
  \def\citenamefont#1{#1}\fi
\expandafter\ifx\csname url\endcsname\relax
  \def\url#1{\texttt{#1}}\fi
\expandafter\ifx\csname urlprefix\endcsname\relax\def\urlprefix{URL }\fi
\providecommand{\bibinfo}[2]{#2}
\providecommand{\eprint}[2][]{\url{#2}}

\bibitem{Zheludev96a} A. Zheludev, S. Maslov, T. Yokoo, J. Akimitsu, S. Raymond, S. E. Nagler,
J. Phys.: Condens. Matter {\bf 13} R525 (2001).

\bibitem{Zheludev98a} A. Zheludev, E. Ressouche, S. Maslov, T. Yokoo, S. Raymond, J. Akimitsu,
Phys. Rev. Lett. {\bf 80}, 3630 (1998); S. Maslov and A. Zheludev,
Phys. Rev. Lett.  {\bf 80}, 5786 (1998); T. Yokoo, S. Raymond, A.
Zheludev, S. Maslov, E. Ressouche, I. Zaliznyak, R. Erwin, M.
Nakamura, and J. Akimitsu, Phys. Rev. B {\bf 58}, 14424 (1998).

\bibitem[{\citenamefont{Alvarez et~al.}(2002)\citenamefont{Alvarez, Valenti,
  and Zheludev}}]{Alvarez02}
\bibinfo{author}{\bibfnamefont{J.~V.}~\bibnamefont{Alvarez}},
  \bibinfo{author}{\bibfnamefont{R.}~\bibnamefont{Valenti}}, \bibnamefont{and}
  \bibinfo{author}{\bibfnamefont{A.}~\bibnamefont{Zheludev}},
  \bibinfo{journal}{Phys. Rev. B} \textbf{\bibinfo{volume}{65}},
  \bibinfo{pages}{184417} (\bibinfo{year}{2002}).


\bibitem[{\citenamefont{Masuda et~al.}(2003{\natexlab{a}})\citenamefont{Masuda,
  Chakoumakos, Nygren, Imai, and Uchinokura}}]{Masuda03a}
\bibinfo{author}{\bibfnamefont{T.}~\bibnamefont{Masuda}},
  \bibinfo{author}{\bibfnamefont{B.~C.} \bibnamefont{Chakoumakos}},
  \bibinfo{author}{\bibfnamefont{C.~L.} \bibnamefont{Nygren}},
  \bibinfo{author}{\bibfnamefont{S.}~\bibnamefont{Imai}}, \bibnamefont{and}
  \bibinfo{author}{\bibfnamefont{K.}~\bibnamefont{Uchinokura}},
  \bibinfo{journal}{J. Solid State Chem.} \textbf{\bibinfo{volume}{176}},
  \bibinfo{pages}{175} (\bibinfo{year}{2003}{\natexlab{a}}).

\bibitem[{\citenamefont{Masuda et~al.}(2004)\citenamefont{Masuda, Zheludev,
  Grenier, Imai, Uchinokura, Ressouche, and Park}}]{Masuda04a}
\bibinfo{author}{\bibfnamefont{T.}~\bibnamefont{Masuda}},
  \bibinfo{author}{\bibfnamefont{A.}~\bibnamefont{Zheludev}},
  \bibinfo{author}{\bibfnamefont{B.}~\bibnamefont{Grenier}},
  \bibinfo{author}{\bibfnamefont{S.}~\bibnamefont{Imai}},
  \bibinfo{author}{\bibfnamefont{K.}~\bibnamefont{Uchinokura}},
  \bibinfo{author}{\bibfnamefont{E.}~\bibnamefont{Ressouche}},
  \bibnamefont{and} \bibinfo{author}{\bibfnamefont{S.}~\bibnamefont{Park}},
  \bibinfo{journal}{Phys. Rev. Lett.} \textbf{\bibinfo{volume}{93}},
  \bibinfo{pages}{077202} (\bibinfo{year}{2004}).

\bibitem[{\citenamefont{Zheludev
  et~al.}(1996{\natexlab{b}})\citenamefont{Zheludev, Shirane, Sasago, Hase, and
  Uchinokura}}]{Zheludev96b}
\bibinfo{author}{\bibfnamefont{A.}~\bibnamefont{Zheludev}},
  \bibinfo{author}{\bibfnamefont{G.}~\bibnamefont{Shirane}},
  \bibinfo{author}{\bibfnamefont{Y.}~\bibnamefont{Sasago}},
  \bibinfo{author}{\bibfnamefont{M.}~\bibnamefont{Hase}}, \bibnamefont{and}
  \bibinfo{author}{\bibfnamefont{K.}~\bibnamefont{Uchinokura}},
  \bibinfo{journal}{Phys. Rev. B} \textbf{\bibinfo{volume}{53}},
  \bibinfo{pages}{11642} (\bibinfo{year}{1996}{\natexlab{b}}).

\bibitem[{\citenamefont{Masuda et~al.}(2003{\natexlab{b}})\citenamefont{Masuda,
  Zheludev, Kageyama, and Vasiliev}}]{Masuda03b}
\bibinfo{author}{\bibfnamefont{T.}~\bibnamefont{Masuda}},
  \bibinfo{author}{\bibfnamefont{A.}~\bibnamefont{Zheludev}},
  \bibinfo{author}{\bibfnamefont{H.}~\bibnamefont{Kageyama}}, \bibnamefont{and}
  \bibinfo{author}{\bibfnamefont{A.}~\bibnamefont{Vasiliev}},
  \bibinfo{journal}{Europhys. Lett.} \textbf{\bibinfo{volume}{63}},
  \bibinfo{pages}{757} (\bibinfo{year}{2003}{\natexlab{b}}).

\bibitem{Lovesey}
{See, for example, S. W. Lovesey, {\it Theory of neutron
scattering from condensed matter vol.2}, (Oxford Science
Publications 1984).}

\bibitem{Masuda05unpublished}
{T. Masuda {\it et al.,} unpublished.}

\bibitem[{\citenamefont{Redhammer and Roth}(2004)}]{Redhammer04a}
\bibinfo{author}{\bibfnamefont{G.~J.} \bibnamefont{Redhammer}}
  \bibnamefont{and} \bibinfo{author}{\bibfnamefont{G.}~\bibnamefont{Roth}},
  \bibinfo{journal}{J. Solid State Chem.} \textbf{\bibinfo{volume}{177}},
  \bibinfo{pages}{2714} (\bibinfo{year}{2004}).

\bibitem[{\citenamefont{Jolicoeur and Golinelli}(1994)}]{Jolicoeur94a}
\bibinfo{author}{\bibfnamefont{Th.}~\bibnamefont{Jolic\oe ur}} \bibnamefont{and}
  \bibinfo{author}{\bibfnamefont{O.}~\bibnamefont{Golinelli}},
  \bibinfo{journal}{Phys. Rev. B} \textbf{\bibinfo{volume}{50}},
  \bibinfo{pages}{9265} (\bibinfo{year}{1994}).

\bibitem[{\citenamefont{Ma et~al.}(1995)\citenamefont{Ma, Reich, Broholm,
  Sternlieb, and Erwin}}]{Ma95a}
\bibinfo{author}{\bibfnamefont{S.}~\bibnamefont{Ma}},
  \bibinfo{author}{\bibfnamefont{D.~H.} \bibnamefont{Reich}},
  \bibinfo{author}{\bibfnamefont{C.}~\bibnamefont{Broholm}},
  \bibinfo{author}{\bibfnamefont{B.~J.} \bibnamefont{Sternlieb}},
  \bibnamefont{and} \bibinfo{author}{\bibfnamefont{R.~W.} \bibnamefont{Erwin}},
  \bibinfo{journal}{Phys. Rev. B} \textbf{\bibinfo{volume}{51}},
  \bibinfo{pages}{R3289} (\bibinfo{year}{1995}).

\end{thebibliography}

\end{document}